\documentclass[apj]{emulateapj}
\usepackage{graphics}

\newcommand{\oiiin}{\mbox{[\ion{O}{3}]}}
\newcommand{\oiii}{\mbox{[\ion{O}{3}]} $\,$}
\newcommand{\oiiiw}{\mbox{[\ion{O}{3}] $\lambda$5007} $\,$}
\newcommand{\oiiiwn}{\mbox{[\ion{O}{3}] $\lambda$5007}}
\newcommand{\hb}{\mbox{H$\beta$} $\,$}
\newcommand{\hbn}{\mbox{H$\beta$}}
\newcommand{\oiiihb}{\mbox{[\ion{O}{3}] $\lambda$5007}/{\mbox{H$\beta$} $\,$}}
\newcommand{\oiiihbn}{\mbox{[\ion{O}{3}] $\lambda$5007}/{\mbox{H$\beta$}}}
\newcommand{\oii}{\mbox{[\ion{O}{2}]} $\,$}
\newcommand{\oiin}{\mbox{[\ion{O}{2}]}}
\newcommand{\nii}{\mbox{[\ion{N}{2}] $\lambda$6583} $\,$}
\newcommand{\niin}{\mbox{[\ion{N}{2}] $\lambda$6583}}

\shortauthors{Comerford et al.}
\shorttitle{Inspiralling Supermassive Black Holes}

\begin{document}

              
\title{Inspiralling Supermassive Black Holes: A New Signpost for
  Galaxy Mergers} 

\author{Julia M. Comerford\altaffilmark{1}, Brian
  F. Gerke\altaffilmark{2}, Jeffrey A. Newman\altaffilmark{3},  Marc
Davis\altaffilmark{1,4}, Renbin
Yan\altaffilmark{5}, Michael C. Cooper\altaffilmark{6,7},
S.M. Faber\altaffilmark{8}, David C. Koo\altaffilmark{8}, Alison
L. Coil\altaffilmark{6,9,10}, D.J. Rosario\altaffilmark{8}, and Aaron
A. Dutton\altaffilmark{8}}   

\affil{$^1$Astronomy Department, 601 Campbell Hall, University of
California, Berkeley, CA 94720} 
\affil{$^2$Kavli Institute for Particle Astrophysics and Cosmology,
  M/S 29, Stanford Linear Accelerator Center, \\ 2575 Sand Hill Rd., 
  Menlo Park, CA 94725}
\affil{$^3$Department of Physics and Astronomy, University of
  Pittsburgh, Pittsburgh, PA 15260}
\affil{$^4$Department of Physics, University of California, Berkeley, CA 94720} 
\affil{$^5$Department of Astronomy and Astrophysics, University of
  Toronto, Toronto, ON M5S3H8, Canada}
\affil{$^6$Steward Observatory, University of Arizona, Tucson, AZ 85721}
\affil{$^7$Spitzer fellow}
\affil{$^8$UCO/Lick Observatory, Department of Astronomy and
  Astrophysics, University of California, Santa Cruz, CA 95064} 
\affil{$^9$Hubble fellow}
\affil{$^{10}$Department of Physics, University of California, San
  Diego, CA 92093}

\begin{abstract}
We present a new technique for observationally identifying galaxy
mergers spectroscopically rather than through host galaxy imaging.
Our technique exploits the dynamics of supermassive black holes
(SMBHs) powering active galactic nuclei (AGNs) in merger-remnant galaxies. 
Because structure in the universe is built up through galaxy mergers
and nearly all galaxies host a central SMBH, some galaxies should possess two SMBHs near their centers as
the result of a recent merger.  These SMBHs spiral
to the center of the resultant merger-remnant galaxy, and one or both
of the SMBHs may   
power AGNs.  Using the DEEP2 Galaxy
Redshift Survey, we have examined 1881 red galaxies, of which
91 exhibit \oiii and \hb emission lines indicative of Seyfert 2 activity.
Of these, 32 AGNs have \oiii emission-line redshifts
significantly different from the redshifts of the host galaxies' stars, corresponding to velocity offsets of 
$\sim
50$ km s$^{-1}$ to $\sim 300$ km s$^{-1}$.  Two of these AGNs exhibit double-peaked
\oiii emission lines, while the remaining 30 AGNs each exhibit a single
set of velocity-offset \oiii emission lines.  
After exploring a variety of physical models for these velocity offsets,
we argue that the most likely explanation is inspiralling SMBHs
in merger-remnant galaxies.
Based on this
interpretation, we find 
that roughly half of the red galaxies
hosting AGNs are also merger remnants, which implies that 
mergers may trigger AGN 
activity in red galaxies. The AGN velocity offsets we find
imply a merger fraction of $\sim 30\%$ and a merger rate of $\sim 3$
mergers Gyr$^{-1}$ for red galaxies at redshifts $0.34 < z < 0.82$. 
\end{abstract}

\section{Introduction }
In the current $\Lambda$ cold dark matter cosmological paradigm, galaxies grow
hierarchically through mergers that combine smaller galaxies to
build a more massive remnant galaxy.  The massive, early-type
galaxies 
observed in the local universe appear to have largely been built up by
such processes from $z \sim 1$ to the present.  The amount of mass in
stars in early-type galaxies has been observed to increase as the
universe evolves to the present time (e.g., \citealt{BU05.2}), which implies
that other galaxies are transforming into early-type galaxies (since
they are not forming stars themselves), and this buildup is believed to be
connected to galaxy mergers (e.g., \citealt{HO07.1}).  Galaxy mergers are thought to be
instrumental not only in the evolution of late-type to early-type galaxies  
\citep{TO72.1}, but also in initiating star formation, triggering inflows of
gas, and initiating galaxy winds that can clear a 
galaxy of its gas \citep{SP05.1}.

Central
supermassive black holes (SMBHs) of a million to a billion solar
masses are found in nearly all galaxies \citep{KO95.1}, and hierarchical structure formation thus implies that some galaxies
should harbor two SMBHs near their centers as the result of a recent
merger. If sufficient gas is available for accretion onto a SMBH one
or both of the SMBHs may power active galactic nuclei (AGNs).

Semianalytic models and numerical simulations show that AGN feedback
quenches star formation and helps transform late-type galaxies
into early-type galaxies \citep{SP05.2,CR06.2}, which implies that we must understand how AGN
activity is initiated in order to understand
galaxy evolution.
Galaxy mergers are in fact thought to play a role in funneling gas
onto galaxy centers and fueling AGNs, as seen in numerical simulations
\citep{SP05.2}.  Determining the fraction of AGNs that are found in
galaxy mergers can help us calibrate the strength of this link.

The frequency of galaxy mergers is also central to our understanding
of galaxy evolution, but the galaxy merger rate has proven
difficult to measure observationally.  The merger rate is usually estimated either from
the number of close dynamical pairs of galaxies (typically defined as
galaxies with separations less 
than 10 -- 40 kpc and radial velocity differences below 500 km s$^{-1}$) or 
the number of galaxies with morphological signatures of mergers such as
tidal features and asymmetries (identified through visual inspection of galaxy images
or quantitative methods).
However, simulations of cosmic structure formation suggest that close
pairs may not be good proxies for mergers \citep{WE08.1} and
that merger timescales are typically underestimated 
\citep{KI08.1}, while
morphological merger identification often misclassifies galaxies 
\citep{DE07.1,LO08.2}.

In this paper, we present a new method of identifying galaxy mergers
that is based on 
inspiralling SMBHs.  Unlike other techniques for identifying galaxy
mergers, our method relies on spectroscopy rather than host galaxy
imaging.  Further, our method involves a very
different set of assumptions than those used in close pair counts or galaxy
morphologies. The complete physical picture of our method is summarized below.

After a merger between two galaxies hosting central SMBHs, dynamical
friction will cause the two SMBHs to inspiral toward the center of
the merger-remnant galaxy.  The two SMBHs will remain at separations
$\gtrsim 1$ kpc for $\sim100$ Myr, then form a SMBH binary of
parsec-scale separation \citep{MI01.1,BE80.1}.  Then, the SMBH binary
must coalesce into a single central  
SMBH in the merger remnant in order to maintain the close observational
correlation between black hole mass and velocity dispersion, or total
mass, of the
host galaxy's stellar bulge \citep{FE00.1}.

If sufficient gas is available, one or both of the SMBHs may
power AGNs.  AGN structure includes a very compact broad-line region
(BLR; $\sim1$ pc in size) in the central part of the AGN and a more extended narrow-line
region (NLR; $\sim100$ pc in size).  If both SMBHs power AGNs,
they could be visible as two
independent AGN nuclei during the $\sim100$ Myr phase of the
merger when their NLRs are not overlapping.  
In this case the
merger-remnant galaxy spectrum would exhibit two sets of AGN
emission lines, including the strong \oiii emission lines that are a
signature of AGN activity.
Both sets of emission lines would be separated both spatially and in
velocity from each  
other and the host galaxy's stellar continuum light. An example of
this type of galaxy is EGSD2 J142033.6+525917, an early-type galaxy
at $z=0.71$ with two sets of Seyfert 2
\oiii lines separated by 630 km s$^{-1}$ and 0.84 $h^{-1}$ kpc \citep{GE07.2}.  We call
such galaxies with two spatially resolved sets of \oiii emission lines ``dual AGNs''.

If only one of the inspiralling SMBHs powers an AGN, then the
merger-remnant galaxy spectrum would display one set of AGN
emission lines that is separated both spatially and in
velocity from the host galaxy's stellar continuum light.  An example
of this type of galaxy is NGC~3341, a
disturbed disk galaxy at $z=0.0271$ with a triple nucleus, where one
nucleus is a confirmed Seyfert 2 with a 
blueshifted velocity of 190 km s$^{-1}$ and a spatial
offset of 5.1 kpc relative to the primary galaxy \citep{BA08.1}.
We call such galaxies with one set of velocity-offset \oiii emission lines ``offset AGNs''.

Because they are the result of inspiralling SMBHs during a galaxy merger, offset and
dual AGNs are a powerful observational tool for identifying galaxy
mergers.  
Offset and dual AGNs are expected to exhibit velocity shifts of up to a few
hundred km s$^{-1}$, and the spatial separations are expected to be $\sim1$ kpc.

While there have been a few serendipitous discoveries of offset and
dual AGNs, in this paper we conduct the first systematic survey of
offset and dual AGNs.  We identify 32 such Seyfert 2 galaxies -- 30
offset AGNs and two dual AGNs -- in the
DEEP2 Galaxy Redshift Survey, and use them to determine both the
fraction of AGNs hosted by 
red galaxy
mergers and the red galaxy merger rate. 
We assume a Hubble constant $H_0 =
100 \, h$ km s$^{-1}$ Mpc$^{-1}$, $\Omega_m=0.3$, and
$\Omega_\Lambda=0.7$ throughout. 

\section{Observations and Analysis}
The recently completed DEEP2 Galaxy Redshift Survey gives the most
detailed view of the $z \sim$ 1 universe currently available.  The
survey covers $\sim 3 \; \mathrm{deg}^2$ of sky, split over four separate
fields, down to a limiting magnitude of $R_{AB}=24.1$. Using
the DEIMOS spectrograph on the Keck II telescope, DEEP2 obtained spectra for 
$\sim$ 50,000 galaxies out to $z=1.4$.  We consider a
sample of 33,211 galaxies that have accurate redshift measurements (quality
$Q=3$ or 4; see \citealt{DA07.3}).

\subsection{Sample Selection}
\label{sampselect}

We first define a subset of galaxies hosting AGNs from the initial
sample of 
33,211 DEEP2 galaxies.  AGN activity can be identified by a host
galaxy's location on the
Baldwin-Phillips-Terlevich (BPT) diagram of line ratios, which is commonly used
to identify the source of line emission in a galaxy \citep{BA81.1,
  KE06.1}. High \oiiihb line flux ratios can indicate
either AGN activity or 
  star formation, and the BPT diagram uses \niin/\mbox{H$\alpha$}
  line flux ratios to distinguish between the two.  High \oiiihb and
  high \niin/\mbox{H$\alpha$} signifies AGN activity, while high \oiiihb
  and low \niin/\mbox{H$\alpha$} signifies star formation. Because the limited wavelength coverage of DEEP2
(6500 -- 9500 \AA) prevents \nii and \mbox{H$\alpha$} from being
covered in the same spectrum as \oiiiw and \hbn, we instead use galaxy
color to distinguish high \oiiihb AGN activity from high \oiiihb star
formation. To accomplish this, we limit ourselves to galaxies on the
red sequence, which necessarily also excludes AGNs that reside in blue galaxies.

To measure the flux of emission lines, we first fit a continuum spectrum to each galaxy spectrum.  We mask out all emission lines present
in each galaxy's spectrum and fit an early-type galaxy template spectrum from
the stellar-population synthesis models of \cite{BR03.1}.  The template
spectrum is a combination of a 0.3 Gyr, solar metallicity, young stellar population and a
7 Gyr, solar metallicity, old stellar population, based on \cite{YA06.1}.

For each galaxy spectrum, we subtract off the continuum and then calculate the flux of the \hb and \oiiiw 
emission lines by summing the one-dimensional spectrum over a 30 \AA $\,$
window ($\sim18$ \AA $\,$ in the rest frame) around each peak.  We
note that our
continuum subtraction accounts for any \hb absorption, enabling an
accurate measurement of the \hb emission-line flux.
We compute rest-frame $\ub$ colors and $M_B$ absolute magnitudes using 
a K-correction code developed for DEEP2 galaxies \citep{WI06.2}.

We make a series of cuts to the full set of DEEP2 galaxies to identify
those hosting AGNs. First, DEEP2 spectra
typically cover the range 6500 -- 9500 \AA, so our sample 
size is limited by the requirement that both \hb and \oiii lines are within
that window.  This restricts us to a redshift range $0.34 < z < 0.82$,
which is outside of the bulk of the DEEP2 sample. Second, to 
ensure that line emission is the result of AGN activity and not
star formation, we require that our sample
has rest-frame $\ub$ colors $\ub > -0.032 (M_B -5 \log h +21.62)+1.285-0.25$,
consistent with a red-sequence galaxy spectrum \citep{WI06.2}. 
These
first two cuts yield a parent sample of 
1881 red galaxies with an absolute magnitude range of $-22.7 < M_B- 5
\log h < -16.8$. 

Next we identify which of these red galaxies
host AGNs. First,  
we require that \oiiiw line emission be detected with at least 3$\sigma$
significance, providing an indication of possible AGN activity.
Second, to ensure that we can robustly measure the \oiiiw emission 
line redshift, we require \oiiiw equivalent width $> 2 \,$
\AA. 
Finally, we require that candidates have line flux
ratios \oiiihb $>3$ to distinguish Seyferts from other line-emitting
galaxies \citep{BA81.1,KA03.3,YA06.1}.
Of the parent sample of 1881 red galaxies, 91 of
them host AGNs as determined by these criteria. We do not limit our
sample by Seyfert type.

\begin{figure}[!t]
\hspace{-.5in}
\begin{center}
\includegraphics[height=3.5in]{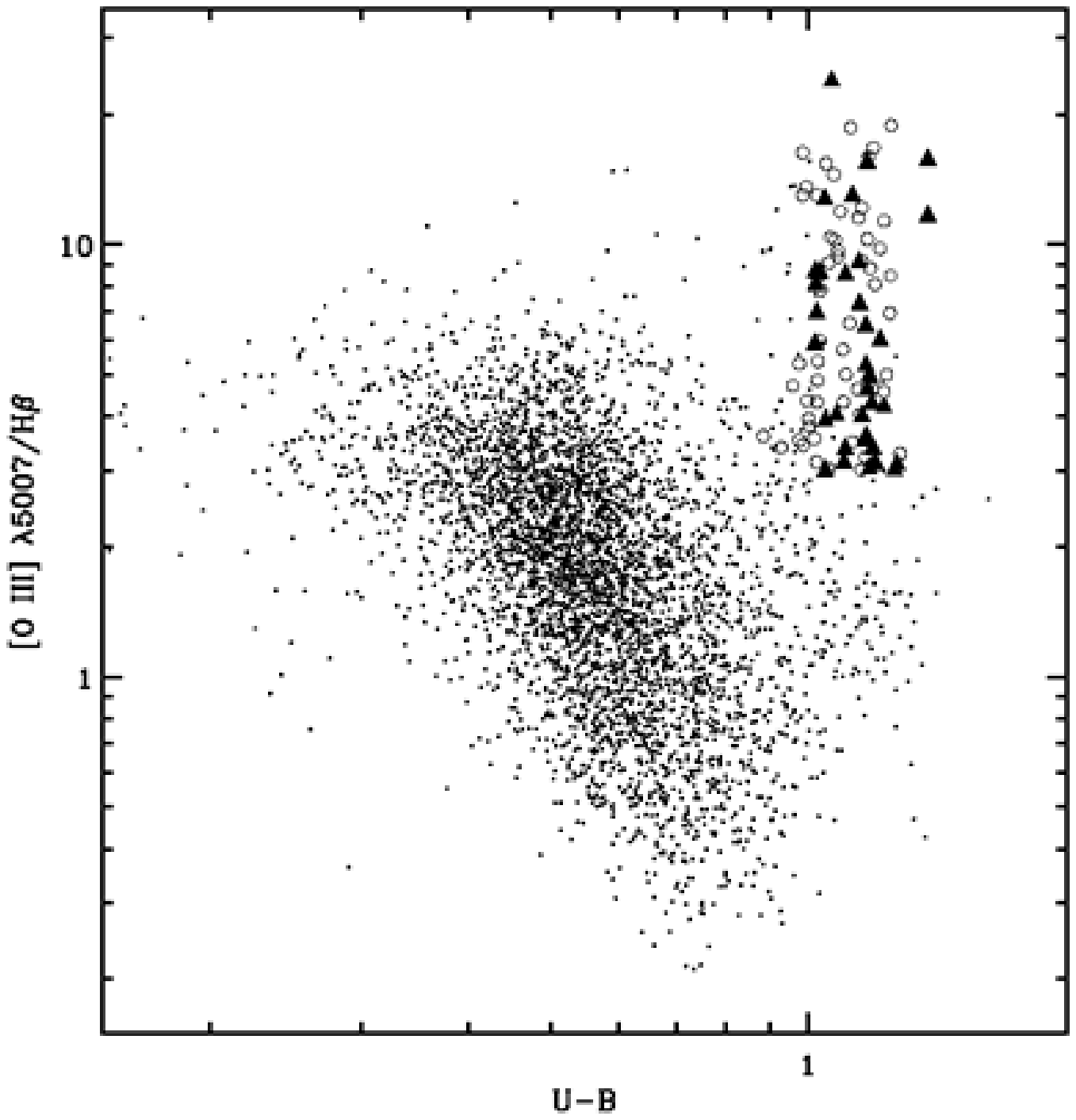}
\end{center}
\caption{\oiiihb line flux ratios and rest-frame $\ub$ colors for our
  sample of DEEP2 galaxies.  Points show the galaxies with accurate
  redshift measurements (quality $Q=3$ or 4; see \citealt{DA07.3}) and
  measurable \oiiihb ratios; open circles show the red galaxies hosting
  AGNs, selected by \oiiihb $>3$ and $\ub > -0.032 (M_B -5 \log h +21.62)+1.285-0.25$
  (note that because of the dependence on absolute magnitude $M_B$,
  this does not correspond to a simple cut in this figure); and filled
  triangles show the 32 offset and dual
  AGNs, which are the red galaxies that host AGNs with
  $>3\sigma$ velocity offsets relative to the host galaxy stars.}
\label{fig:selection}
\end{figure}

Figure~\ref{fig:selection} illustrates the differences between the
population of red galaxies hosting AGNs and the general population of
galaxies.  Galaxies populate this diagram of \oiiihb versus $\ub$ similar
to the way they populate the traditional BPT diagram of
\niin/\mbox{H$\alpha$} versus \oiiihbn, disassociating into a star-forming
sequence and an AGN plume.  The separation between these populations
is less distinct here than in the BPT diagram, but the figure
illustrates that we have selected a well-defined sample of galaxies
that are well separated from the main star-forming sequence. 

\subsection{Measurement of AGN Velocity Offsets}
\label{measure}

The line width of AGN \oiii emission usually correlates with the
velocity dispersion of the stellar bulge of the host galaxy, 
implying that the \oiii emission component is generally at rest with
respect to the stellar bulge and, by extension, the central SMBH
\citep{GR05.1}.  In that case, the \oiii emission lines should be at
the same redshift as the stellar component of the host galaxy.
If two galaxies with central SMBHs have recently undergone a merger,
however, the merger
remnant should host two SMBHs moving relative to one another.  If one
or both of the SMBHs are active, we expect the AGN \oiii
emission lines to be at a different redshift than the remnant's stars,
whose mean velocity will be in the rest frame of the remnant.

To identify such cases, we first measure the galaxy redshift based on the wavelengths of the diversity of absorption
features in the stellar spectrum.  We accomplish this by fitting
template spectra to the observed spectra, as described in Section~\ref{sampselect}.
We then determine the redshift that yields 
minimum $\chi^2$ for the fit of the galaxy template spectrum to the
actual spectrum. In performing this fit, we include the
  variance in the  
  one-dimensional spectrum at each wavelength position.  We compute
  the variance by propagating
  errors from the Poisson uncertainty in photon counts on each
  pixel of the detector, including the effects of night-sky background
  and cosmic rays. Our sky subtraction does not  
  introduce significant systematic errors beyond this simple Poisson
  noise (Faber et al., in preparation).
  We determine the uncertainty in the absorption redshift as the
width of the $\chi^2$ minimum at which $\Delta \chi^2=1$.

Then, we measure the emission redshift using a robust two-step
centroiding algorithm.  First, we take a  30 \AA $\,$ window ($\sim18$
\AA $\,$ in the rest frame) around
each galaxy's \oiiiw emission line and fit a Gaussian to the line.
We use the width of this Gaussian to define a narrow window centered
on the peak of the emission line, within which we compute the line
centroid.  This procedure gives a centroid measurement whose value is
robust to noise in the outer wings of the line. Using Monte Carlo
realizations 
drawn from the variance in the spectra, we find that our centroid
values 
are accurate to better than 0.1 \AA $\,$ for typical signal-to-noise
ratios. 

We then verify each galaxy's absorption and emission-line
redshift by eye, and in the process reject the $5 \%$ of objects in
which the fit appears to be dominated by noise and the signal-to-noise
ratio is too low to accurately determine an 
absorption redshift. 

A discrepancy between absorption and emission redshifts could indicate that
the \oiii emission component (the AGN) is moving with respect to the
absorption component (the host galaxy's stars). We convert the redshift
difference to a radial velocity separation in the host's
rest frame, and we derive the error on this velocity separation from
the errors on the two redshifts. With these velocity separations,
we identify offset and dual AGNs. 

\begin{deluxetable*}{lccclll}
\tablewidth{0pt}
\tablecolumns{7}
\tablecaption{Host Galaxy Properties of the 32 Offset and Dual AGNs} 
\tablehead{
\colhead{ID} &
\colhead{$\ub$} &
\colhead{$M_B$} &
\colhead{{\mbox{[\ion{O}{3}] $\lambda$5007}/}} &
\colhead{$z_{abs}$} & 
\colhead{$z_{em}$} &
\colhead{$v_{em}-v_{abs}$} \\ 
& 
& 
\colhead{$-5 \log h$} &
{\mbox{H$\beta$}} &
&
&
\colhead{(km s$^{-1}$)} 
}
\startdata
EGSD2 J141515.6+520354 & 1.04 & -20.9 & 12.9 & 0.69304 &
0.69355 & \phantom{-}\phantom{0}89.9 $\pm$ 14.6 \\
EGSD2 J141417.6+520351 & 1.10 & -20.5 & 3.18 & 0.77447 &
0.77404 & \phantom{0}-72.9 $\pm$ 17.7 \\
EGSD2 J141523.5+520532 & 1.20 & -20.3 & 3.16 & 0.48214 &
0.48265 & \phantom{-}103.9 $\pm$ 16.9 \\
EGSD2 J141550.8+520929 & 1.17 & -20.6 & 4.69 & 0.61987 &
0.61935 & \phantom{0}-97.0 $\pm$ 19.8 \\
 &  &  & 3.58 &  & 0.62171 & \phantom{-}339.5 $\pm$ 19.8 \\
EGSD2 J141547.7+520843 & 1.08 & -20.9 & 4.09 & 0.61916 &
0.61806 & -203.0 $\pm$ 20.6 \\
EGSD2 J141458.0+520915 & 1.18 & -19.3 & 3.08 & 0.48358 &
0.48396 & \phantom{-}\phantom{0}76.1 $\pm$ 19.6 \\
EGSD2 J141433.1+520834 & 1.17 & -20.9 & 5.34 & 0.77156 &
0.77261 & \phantom{-}178.7 $\pm$ 22.6 \\
EGSD2 J141643.2+521721 & 1.26 & -20.4 & 3.06 & 0.45095 &
0.45133 & \phantom{-}\phantom{0}77.0 $\pm$ 11.5 \\
EGSD2 J141732.6+523817 & 1.19 & -22.0 & 3.43 & 0.71671 &
0.71792 & \phantom{-}211.5 $\pm$ 12.1 \\
EGSD2 J141711.0+523729 & 1.18 & -21.1 & 4.36 & 0.64395 &     
0.64344 & \phantom{0}-92.1 $\pm$ 20.3 \\
EGSD2 J141839.2+525140 & 1.17 & -20.3 & 3.65 & 0.34514 &
0.34546 & \phantom{-}\phantom{0}71.4 $\pm$ 18.4 \\
EGSD2 J142017.9+525538 & 1.02 & -18.7 & 5.98 & 0.63750 &
0.63843 & \phantom{-}169.9 $\pm$ 30.0 \\
EGSD2 J142043.0+525716 & 1.18 & -19.7 & 5.06 & 0.74898 &
0.74831 & -115.0 $\pm$ 24.9 \\
EGSD2 J142033.6+525917 & 1.38 & -20.2 & 11.8 & 0.70882 &     
0.70701 & -318.3 $\pm$ 34.1 \\
 & & & 15.9 & & 0.71060 & \phantom{-}311.0 $\pm$ 34.1 \\
EGSD2 J141939.0+530223 & 1.05 & -20.2 & 3.04 & 0.76309 &
0.76243 & -111.4 $\pm$ 30.4 \\
EGSD2 J141929.7+530104 & 1.10 & -21.7 & 3.42 & 0.67789 &
0.67727 & -110.9 $\pm$ 13.2 \\
EGSD2 J142010.1+530738 & 1.15 & -20.4 & 4.07 & 0.57380 &
0.57468 & \phantom{-}167.6 $\pm$ 32.7 \\
EGSD2 J142153.6+531352 & 1.02 & -20.7 & 8.20 & 0.67257 &
0.67205 & \phantom{0}-93.9 $\pm$ 18.3 \\
EGSD2 J142057.2+531104 & 1.21 & -20.0 & 6.09 & 0.71417 &
0.71372 & \phantom{0}-78.8 $\pm$ 17.3 \\
EGSD2 J142143.0+531820 & 1.17 & -20.8 & 6.58 & 0.76479 &
0.76501 & \phantom{-}\phantom{0}37.1 $\pm$ 12.3 \\
DEEP2 J164714.9+350405 & 1.22 & -20.3 & 4.28 & 0.76186 & 0.76147 & \phantom{0}-65.7
$\pm$ 14.7 \\
DEEP2 J164646.6+350648 & 1.02 & -20.8 & 8.82 & 0.74588 & 0.74629 & \phantom{-}\phantom{0}69.2
$\pm$ 13.5 \\
DEEP2 J165128.8+344841 & 1.03 & -20.8 & 8.75 & 0.70433 & 0.70381 & \phantom{0}-90.7
$\pm$ 16.5 \\
DEEP2 J232717.4+000803 & 1.15 & -20.4 & $\hspace{-.9mm} >$9.24\tablenotemark{a} & 0.74173 & 0.74142 & \phantom{0}-54.3
$\pm$ 16.9 \\
DEEP2 J232907.8+001742 & 1.06 & -21.4 & 24.3 & 0.79166 & 0.79137 & \phantom{0}-47.8
$\pm$ 6.3 \\
DEEP2 J233030.0+002418 & 1.02 & -21.0 & 7.05 & 0.77831 & 0.77922 & \phantom{-}153.7
$\pm$ 15.6 \\
DEEP2 J233250.1+001929 & 1.11 & -20.2 & 8.65 & 0.70228 & 0.70291 & \phantom{-}110.1
$\pm$ 22.4 \\
DEEP2 J022735.0+003816 & 1.13 & -20.5 & 13.1 & 0.67491 & 0.67449 & \phantom{0}-75.7
$\pm$ 16.2 \\
DEEP2 J023050.3+002408 & 1.15 & -20.6 & 7.39 & 0.62009 & 0.62063 & \phantom{-}100.4
$\pm$ 16.5 \\
DEEP2 J023059.6+004418 & 1.05 & -20.1 & 3.97 & 0.77672 & 0.77618 & \phantom{0}-90.4
$\pm$ 20.6 \\
DEEP2 J022907.1+004353 & 1.17 & -21.3 & 15.7 & 0.65378 & 0.65299 & -143.0
$\pm$ 16.0 \\
DEEP2 J023121.3+005110 & 1.27 & -22.7 & 3.15 & 0.77503 & 0.77665 & \phantom{-}273.9
$\pm$ 8.5 \\
\enddata

\tablenotetext{a}{For this object, the measurement of \hb flux is
  consistent with zero, so we use the 2$\sigma$ upper limit on
  \hb to derive a lower limit for \oiiihbn.}
\label{tbl:agn}
\end{deluxetable*}

\subsection{32 Offset and Dual AGNs}
\label{32}

To avoid contamination from objects with nonzero velocity differences
due to measurement errors, we restrict our offset AGN sample to 
only those with velocities different from zero by more than 3$\sigma$, 
which excludes all cases with differences $\lesssim 50$ km s$^{-1}$.
We are sensitive to relatively small velocity differences due to the
high spectral resolution ($R \sim 5000$) and excellent sky
subtraction of DEEP2 data. Absorption-line redshift uncertainties
estimated from the width of the $\chi^2$ minimum (described in
Section~\ref{measure}) are 
small: 6 -- 34 km s$^{-1}$ for our sample, with a median of 17 km s$^{-1}$.
Uncertainties in the \oiii emission-line redshifts (described in
Section~\ref{measure}) are even
smaller, $\lesssim$ 1 km s$^{-1}$.  These errors are statistical only; they
do not include systematic errors that arise from, for example, a
galaxy not being perfectly centered in its slit. 
However, to first order position errors will have an identical effect
on the observed velocity of a galaxy and any AGN it harbors.

As a check on our error estimates, we have compared the differences
between the AGN emission redshift and the host galaxy absorption redshift
for the three AGN host galaxies observed  
twice by DEEP2; the different observations yield velocity offsets
differing by 4.2 km s$^{-1}$, 9.4 km s$^{-1}$, and 15.2 km s$^{-1}$.  Hence, we can 
find no evidence that our redshift uncertainties have been
significantly underestimated. 

After we reject the 59 galaxies 
with measured velocity differences below $3 \sigma$ significance, the final
sample consists of 30 offset AGNs and two dual AGNs, for a total of 32
objects. This number is a lower bound on the true number of
offset and dual AGNs in the sample, because we
  exclude low-significance velocity offsets and are also not sensitive to
  velocity components perpendicular to the line of sight.  We do not select by
Seyfert type, 
but all 32 offset and dual AGNs are in Seyfert 2 galaxies.

Figure~\ref{fig:selection} shows that the offset and dual AGNs do not
have systematically different \oiiihb ratios or colors from the
general population of AGNs, but rather are uniformly distributed.
Apart from their velocity offsets, the offset and 
dual AGNs appear to be typical Seyferts.
Table~\ref{tbl:agn} gives the rest-frame $\ub$ colors, absolute
magnitudes, \oiiihb ratios, absorption and emission redshifts, and
velocity offsets for each of the 32 offset and dual AGNs.

\begin{figure}[!t]
\hspace{-.5in}
\begin{center}
\includegraphics[height=3.5in]{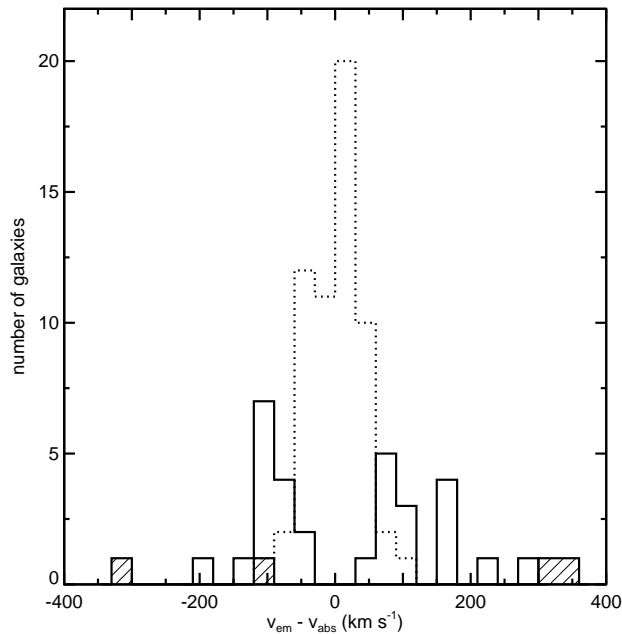}
\end{center}
\caption{Differences between the velocity of AGN emission lines ($v_{em}$) and
  the velocity of the host galaxy's stars ($v_{abs}$) for the full
  sample of 91 AGNs. The 
  open histogram depicts the velocity differences of the 30 offset AGNs,
  the hatched histogram depicts the velocity differences for each
  component of the two 
  dual AGNs, and the dotted histogram depicts the 59 objects with less
  than $3 \sigma$ velocity 
  difference that were removed from our sample to
  eliminate contamination due to measurement errors
  from objects with no 
  statistically significant velocity difference. 
  The velocity-difference
  distribution of the remaining AGNs is symmetric, indicating
  that an offset AGN emission line is 
  equally likely to be redshifted or blueshifted
  relative to the rest frame of the stellar continuum.}
\label{fig:vhist}
\end{figure}

Figure~\ref{fig:vhist} shows a histogram of the difference
between the velocity of the emission component (the AGN) and the
absorption component (the host galaxy's stars) for the 32 offset and
dual AGNs.  
Because the histogram is symmetric about zero velocity
difference, we see no evidence for bias in the radial direction of the
velocity of the emission-line region. The dotted histogram shows the 59 AGNs that exhibit less
than $3 \sigma$ velocity differences, preventing their classification
as offset AGNs.

\begin{figure}[!t]
\centering
\hspace{-.39in}
\includegraphics[scale=.5]{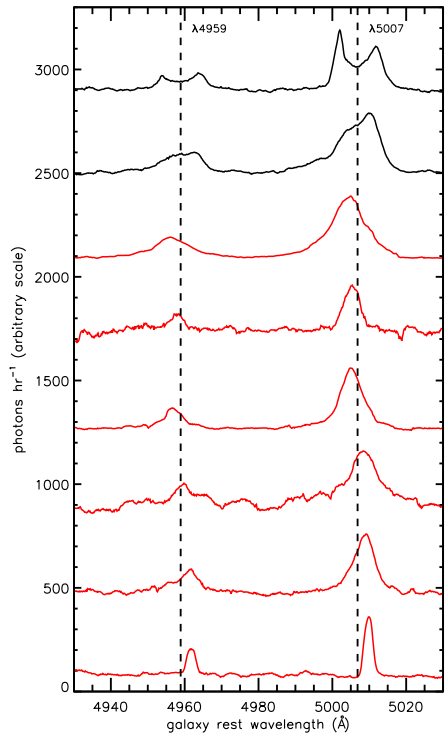}
\caption{Segments of the one-dimensional DEIMOS spectra of the host
  galaxies of the two dual AGNs (shown as the top two, in black) and
  six typical offset AGNs (shown as the bottom six, in red). For
  clarity, the spectra are offset from one another vertically and
  normalized to span 300 counts hr$^{-1}$. Each 
  spectrum is shifted to the rest frame of the host galaxy's stars,
  weighted by its inverse variance, and smoothed by a smoothing
  length of 1.5 \AA.  The dashed lines show the expected wavelengths
  of \oiii at 4959 \AA $\,$ and 5007 \AA. 
  From top to bottom, the
  spectra shown are of EGSD2 J142033.6+525917, EGSD2 J141550.8+520929,
  DEEP2 J022907.1+004353, EGSD2 J142043.0+525716, EGSD2
  J142153.6+531352, EGSD2 J141515.6+520354, DEEP2 J233250.1+001929,
  and EGSD2 J142017.9+525538. From top to bottom, the
  velocity separations between the two \oiii emission line peaks in the
  dual AGNs are  
  630 km s$^{-1}$ and 440 km s$^{-1}$, while the velocity offsets for the \oiii emission 
  lines in the offset AGNs are -140
  km s$^{-1}$, -120 km s$^{-1}$, -94 km s$^{-1}$, 90 km s$^{-1}$, 110
  km s$^{-1}$, and 170 km s$^{-1}$.}
\label{fig:offsets}
\end{figure}

Figure~\ref{fig:offsets} depicts example offset and dual AGNs plotted
in the rest frame of the stellar continuum, where the dashed lines
show the
expected locations of the \mbox{[\ion{O}{3}]} $\lambda\lambda$ 4959,
5007 emission lines in that rest frame.
The bottom six spectra (shown in red) are
example spectra for 
six offset AGNs, where for clarity we set off each spectrum
vertically from the others. The plot shows AGNs that are both blueshifted and
redshifted with respect to their host galaxies, ranging from a
blueshift of -143 km s$^{-1}$ to a redshift of 170 km s$^{-1}$. 

\begin{figure*}[!t]
\hspace{-.1in}
\begin{center}
\includegraphics[scale=.6]{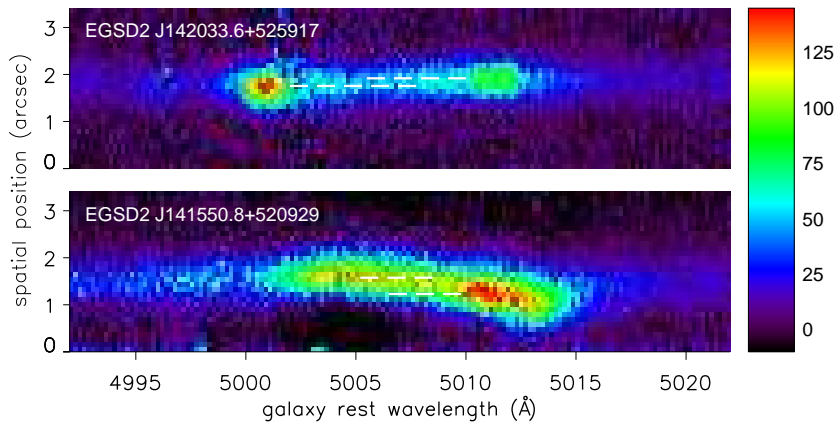}
\end{center}
\caption{\oiiiw emission in the two-dimensional DEIMOS spectra of the two
  dual AGNs, with night-sky emission features subtracted.
  The top panel depicts EGSD2 J142033.6+525917 at
  $z=0.71$, and the bottom panel depicts EGSD2 J141550.8+520929 at
  $z=0.62$. 
  In both
  panels, the vertical axis spans 32 DEIMOS pixels (3$^{\prime\prime}$.5) in spatial position along the slit
  and the horizontal axis spans 30 \AA $\,$ in rest-frame wavelength
  centered on
  \oiiiwn.  The horizontal, white dashed lines show the spatial
  centroids of each emission-line component.  The two emission-line
  components in the top panel are separated by 1.5 DEIMOS pixels,
  which is equivalent to 0$^{\prime\prime}$.17 or 0.84 $h^{-1}$ kpc, and the two
  emission-line components in the bottom panel are separated by 3.1
  DEIMOS pixels, which is equivalent to 0$^{\prime\prime}$.34 or 1.6 $h^{-1}$ kpc. 
The slanted,
  nearly-vertical features are imperfectly subtracted night-sky lines,
  and the color bar provides a scale for flux in counts/hour/pixel.}
\label{fig:dualAGN}
\end{figure*}

The top two spectra (shown in black) in Figure~\ref{fig:offsets} depict the
two dual AGNs in our sample. The multislit spectroscopy used in the
DEEP2 survey also enables us to make two-dimensional spectra of
spatial position and wavelength for each galaxy.  The slits are 1$^{\prime\prime}$
wide and vary in length from typically $\sim5^{\prime\prime}$ to $\sim10^{\prime\prime}$. 
Figure~\ref{fig:dualAGN} shows the
two-dimensional DEEP2 spectrum centered around each object's \oiiiw
emission. Each dual AGN has double-peaked emission at \oiiiwn, and we
determine the spatial centroid of each emission component by applying the
technique described in Section~\ref{measure} for measuring the emission
redshift, only we apply this technique along the spatial direction of
the two-dimensional spectrum rather than
the wavelength direction. 

The dual AGN EGSD2 J142033.6+525917
at redshift $z=0.71$ is shown at the top of Figure~\ref{fig:dualAGN}.  Its spectrum features double-peaked
\oiii emission lines
separated by 630 km s$^{-1}$ in the host rest frame \citep{GE07.2}.
From the 
two-dimensional DEIMOS spectrum of the object, the projected physical
separation between the pair of emission peaks is 1.5 DEIMOS pixels, or
0$^{\prime\prime}$.17, which corresponds to a projected 
physical separation of 0.84 $h^{-1}$ kpc. 

The dual AGN
EGSD2 J141550.8+520929 at $z=0.62$ is shown at the bottom of
Figure~\ref{fig:dualAGN}.  Its
two-dimensional DEEP2 spectrum shows two overlapping \oiii
emission lines.  The velocity 
separation between the two emission components is 440 km s$^{-1}$ and
the spatial offset of their centroids is 3.1 DEIMOS pixels, or 0$^{\prime\prime}$.34,
which corresponds to a projected 
physical separation of 1.6 $h^{-1}$ kpc. 

\subsection{Extrapolated Number of Low-Velocity-Offset AGNs}
\label{low}

By excluding objects with less than $3\sigma$ significance, we will discard
many AGNs with small, but real, velocity offsets. 
We can measure only the radial component of a velocity, which is a
factor of sin$\, i$ less than the total velocity at an angle of
inclination $i$ relative to the perpendicular to the line of sight. 
Because of these projection effects, we expect an even greater abundance of
small measured velocity offsets than large measured offsets.  If we fit a
Gaussian centered at zero to the offset AGN velocity distribution shown in Figure~\ref{fig:vhist}, we find that
23 objects with absolute velocity offsets $<$ 50 km s$^{-1}$ should be added to the
sample to create a Gaussian distribution of velocity offsets.  

More conservatively, we can assume that the velocity-offset
distribution is flat in the central region around zero.  If we assume
the number of offset AGNs with absolute velocity offsets $<$ 50 km s$^{-1}$ is
determined by the mean number of offset AGNs with absolute offsets ranging from
50 -- 75 km s$^{-1}$, this would instead increase the sample size by eight
objects with $<$ 50 km s$^{-1}$ absolute velocity offsets.

We add 23, or more conservatively eight, low-velocity-offset AGNs to the 32 definitively detected offset and dual AGNs.  As a result, we expect 40 -- 55 offset and dual AGNs in our full sample of 91 AGNs.

\section{Interpretation of Offsets}
\label{interpret}

We have measured statistically significant velocity offsets of \oiii
emission lines in 32 offset and dual AGNs in DEEP2 red galaxies.  
Here, we explore the physical mechanisms that could cause the \oiii
velocity offsets we observe: small-scale gas kinematics, AGN outflows
and jets, recoiling SMBHs due to gravitational 
radiation emission after SMBH coalescence, and inspiralling SMBHs in a galaxy merger.  We find that the most plausible explanation for our offset and dual AGNs is inspiralling SMBHs in a galaxy merger remnant.

\subsection{Small-scale Gas Kinematics}
\label{gas}

Gas rotation and bulk flows on small scales can produce velocity shifts
in \oiii emission lines.  Here we review observations of \oiii
velocity shifts that can be explained by small-scale gas kinematics
and examine whether such effects could explain the velocity offsets in
our sample.

In an analysis of 13 Seyfert galaxies
and quasi-stellar objects (QSOs) at $0.003 < z < 0.04$, \cite{VR85.1} measured \oiii
emission-line velocity offsets ranging from 0 -- 180 km s$^{-1}$. Five
of the 13 galaxies have \oiii velocity offsets greater than 3$\sigma$
in significance, which is our definition of an offset AGN (Section~\ref{32}).
Of the five Seyferts with statistically significant \oiii velocity offsets,
three exhibit redshifted \oiii and two exhibit blueshifted \oiiin.
These observations were taken with an echelle spectrograph.  If the
slits were slightly miscentered on the galaxy centers, then rotation of gas in disks of order 100 pc in size or small-scale bulk flows of
gas near the galaxy centers could produce the \oiii velocity offsets
measured by \cite{VR85.1}.

In
addition, observations of 54 Seyferts at $0.002 < z < 0.04$ revealed \oiii
emission-line velocity offsets of 1 -- 230 km s$^{-1}$
\citep{NE95.1}.  25/54 of the velocity offsets have greater than
3$\sigma$ significance, and of these six exhibit redshifted \oiii and 19
exhibit blueshifted \oiiin.  These velocity offsets could be
explained by slit miscentering and small-scale gas kinematics, as
described above. However, because there are a factor of 3 more blueshifts
than redshifts, AGN outflows may be the likelier explanation
(Section~\ref{outflow}). 

For completeness, we also note that although some analyses of double-peaked broad lines in
quasar spectra conclude they are caused by binary black
holes (e.g.,
\citealt{GA84.1,GA88.1,GA96.1}), most such lines are now understood to
be caused by accretion discs (e.g., \citealt{ER97.2,ER03.2}).  However,
these are a very 
different class of objects than our sample of double-peaked AGNs.  The
quasar sample
exhibits emission from the BLR on $\sim$1 pc scales that produces double-peaked broad lines with line widths comparable to their
velocity separations, whereas our dual AGNs exhibit emission from the
NLR on $\sim100$ pc scales that produce
double-peaked narrow lines with line widths smaller than their
velocity separations.  Because we probe different sets of lines that
arise from much larger scales, the interpretations of the
double-peaked broad lines do not 
necessarily transfer to our sample of dual AGNs.

Though they are a possible explanation for the \cite{VR85.1} and
\cite{NE95.1} velocity offsets and the double-peaked broad line quasar
spectra, small-scale gas rotation or
small-scale bulk flows of gas alone cannot explain our observations of \oiii
velocity offsets. Our observations are at the much higher redshift
range $0.34 < z < 0.82$, where our 1$^{\prime\prime}$
slit width subtends several kpc across each galaxy.  As a result, our
observations are based on a
full spatial average of the NLR kinematics, which means that
small-scale gas kinematics alone cannot cause the \oiii velocity
offsets we measure.

If the small-scale gas kinematics were accompanied by local dust, one can
imagine scenarios where only the blueshifted (redshifted) gas is
obscured, leaving 
the redshifted (blueshifted) gas to manifest as velocity-offset
emission lines. However, for this scenario to occur the local dust would
need to be pathologically patchy rather than in the usual form
of a ring or torus. Dust on larger scales, such as in a disk or lane
on kpc or 10 kpc scales, would blot out the entire central region and
result in no velocity-offset detection. 

While small-scale gas kinematics can lead to observations of \oiii
velocity offsets at low redshifts, they cannot alone produce the
velocity offsets we measure at higher redshifts.  A
combination of small-scale gas kinematics and patchy dust could
explain our velocity offsets, though the dust must be in an unusual
configuration. 

\subsection{AGN Outflows and Jets}
\label{outflow}

Velocity shifts of \oiii emission lines can also be explained by a strong,
decelerating wind in the inner 
NLR of the AGN \citep{KO08.1}.  For a galactic wind,
unlike a stellar wind, the mass enclosed increases as the galactic wind
flows outward, and causes the galactic wind to decelerate with distance
from the galactic center.
Studies of outflows suggest that the NLR around the AGN is stratified such that high-ionization lines, such as \oiiin, are preferentially generated near the
AGN and low-ionization lines, such as \hb and \oiin, are preferentially 
generated further from the AGN (e.g., \citealt{ZA02.1,KO08.1}).  
These
outflow studies find that a centrally 
driven outflow that decelerates with distance from the AGN would
impart the highest velocities to the nearby high-ionization lines, and
low-ionization lines further from the AGN would exhibit lower velocity
offsets or remain stationary with respect to the galaxy's stellar
component.   

If the AGN has a bulk motion within the host galaxy,
however, then all AGN emission lines, regardless of ionization
potential, would exhibit the same velocity shifts relative to the host
galaxy's stars.  While only nine galaxies in our sample have \oii lines within
our wavelength window of observation and 31 have sufficiently high \hb
signal-to-noise ratios
to accurately determine a redshift, in galaxies where the measurements
are possible we find
that \mbox{[\ion{O}{2}]}, \mbox{H$\beta$}, and \oiii lines have consistent
velocity offsets to within 1$\sigma$.  These AGNs all exhibit bulk motions not consistent
with the stratified velocity structure of outflows.

We also compare to other samples of AGNs exhibiting velocity offsets
due to outflows to
understand the physical mechanism behind our offsets.
For a sample of $\sim 200$ quasars at $z
<0.8$, \cite{ZA02.1} measured the \oiiiw redshift relative to
the \mbox{H$\beta$} redshift, taken to be the systemic redshift of
the galaxy. In such luminous objects, stellar
absorption features are overwhelmed by light from the quasar.
As a result, the host galaxy redshift cannot be derived from the
stellar absorption lines as we do with the DEEP2 sample.
Of the objects with inconsistent \oiiiw and
\mbox{H$\beta$} redshifts, more had blueshifted \oiiiw than 
redshifted \oiiiwn, relative to \mbox{H$\beta$}.  This includes a factor of
3 more blueshifts of radial velocity $v_r$ in the range -200 km s$^{-1}$
$\lesssim \Delta v_r \lesssim$ -100 
km s$^{-1}$ than redshifts of 100 km s$^{-1}$ $\lesssim \Delta v_r \lesssim$ 200 km s$^{-1}$.
AGN outflows are expected to exhibit blueshifts, as
material traveling away from the observer will be on the far side of the AGN
and obscured by its dust torus. 
The outliers with blueshifted \oiiiw also extended out to $\Delta v_r
\approx$ -1000 km s$^{-1}$, while no redshifted \oiiiw was observed beyond $\Delta v_r
\approx$ 280 km s$^{-1}$. Because the distribution of velocity offsets is
skewed toward the blue, \cite{ZA02.1} concluded that outflows in the
inner NLR of the AGN influence the peak velocity of
the \oiiiw line.  The \cite{NE95.1} sample also has a velocity distribution skewed
toward the blue, suggesting that outflows may be the root of those velocity
offsets.

In our sample, however, there are 
equal numbers of blueshifted and redshifted outliers (15 of each)
and both blueshifted and redshifted outlier
distributions extend to similar velocity offsets ($\Delta v_r$=-203
km s$^{-1}$ and $\Delta v_r$=274 km s$^{-1}$, respectively). Because the
distribution of \oiiiw velocity offsets in our sample is quite
different from that of \cite{ZA02.1}, the physical
explanations for the 
offsets in the two samples are likely different.

Furthermore, outflows are generally found to cause a strong correlation of
increasing \oiii line width with \oiii blueshift (e.g., \citealt{KO08.1}). 
By contrast, our sample has a correlation coefficient of 0.009,  
indicating a very weak correlation between \oiii line width
and \oiii velocity shift.  

To explore how small-scale outflows would affect
observations of our 
high-redshift Seyfert galaxies, we examine
nearby Seyfert galaxies with outflows on scales of 100 pc.
Using the {\it Hubble Space Telescope} Imaging
Spectrograph (STIS), \cite{DA06.1} obtained spatially resolved 
spectroscopy of outflowing gas in the NLR of the Seyfert 2 galaxy NGC~1068.  The outflowing clouds produce a complicated velocity structure
in the \oiii emission lines, with multiple emission components having
velocity offsets (both blueshifts and redshifts) up to several hundred
km s$^{-1}$.  This velocity structure is confined to the inner $\sim
100$ pc of the emitting region; outside of this, the outflow velocity
drops and reaches zero at $\sim 300$ pc.  (Similar kinematics and spatial structure are
seen in the Seyfert 1 galaxy NGC~4151; \citealt{DA05.2}.)  By
comparison, a DEEP2 slit of $1^{\prime\prime}$ width subtends
approximately 7 kpc at $z \sim 0.7$ (where most of our sample lies)
and a DEIMOS pixel 
subtends approximately 1 kpc, so similar kinematics on subkiloparsec
scales would not be resolved in our observations.  To get a rough
approximation of what the spectrum of a system like NGC~1068 would look
like in DEEP2, we can coadd the neighboring STIS spectra of NGC~1068
from \cite{DA06.1}; this yields a broadened \oiii emission line
with a blue wing but no
large overall offset in the velocity of its peak.

Another example of small-scale outflows is the \cite{RU05.1} sample,
which consists of STIS observations of 10 Seyfert galaxies at $0.003 <
z < 0.03$.  Six of the 10 objects have observable \oiii velocity
offsets, ranging from 70 -- 200 km s$^{-1}$.  The disturbed kinematics
of these galaxies result in \oiii FWHMs that are generally larger than
the \oiii velocity offsets, and such increases in \oiii FWHM may also
be the result of radio jets driving the outflows \citep{NE96.1}.

Our sample of offset and dual AGNs does not exhibit this trend of larger \oiii FWHMs
than velocity offsets, and our approximation of NGC~1068 at a higher
redshift suggests that we would not detect a velocity offset in such an object.
As a result, the nearby systems we examine show no evidence that typical small-scale Seyfert outflows would lead to the
kind of emission-line offsets we observe in our sample.

To understand the effect of large-scale outflows on observations of
our sample of Seyferts, we consider nearby Seyfert galaxies with
outflows on scales of 10 kpc. These systems exhibit disturbed
kinematics, with \oiii velocity dispersions that are larger than the
\oiii velocity offsets (e.g., \citealt{VE01.1, GE09.1}), but this
effect is not seen in our offset and dual AGNs.  Also, in the local
systems the outflowing regions tend to be much fainter than the
stationary central NLR.  In a composite spectrum of such an object at
the redshift of our sample, the \oiii flux would be dominated by the
brighter, stationary central component and we might not detect a
velocity offset. 

Another scenario is that dust in the galaxy, in concert with an
outflow, could result in a detection of an \oiii velocity offset.
As described above, a local dust torus would result in an excess of
\oiii blueshifts rather than our sample's symmetric distribution of
\oiii redshifts and blueshifts.  Dust on larger scales that obscures
only a portion of the outflow could result in the observation of an
\oiii velocity offset, but such configurations of dust have not been
seen locally.  Further, our sample consists of red
galaxies that should not have much dust, and AGNs that are luminous
enough to drive large-scale outflows would also likely expel dust.
Even if a conspiracy between dust and large-scale outflows led to
observations of \oiii velocity offsets, the \oiii velocity dispersions
would still be larger than the velocity offsets, which is not the case
in our sample.

Our above comparisons to local galaxies with small-scale or
large-scale outflows suggest that the \oiii velocity offsets in
our sample are not consistent with the outflow explanation.  However,
a combination of partial dust obscuration and outflow activity, however
unlikely, could explain the velocity offsets we observe.

\subsection{Recoiling SMBHs}

Another possible interpretation of our offset AGNs is that some may be
recoiling SMBHs. 
After a galaxy merger two SMBHs can coalesce due to
gravitational wave emission, which can carry away linear momentum and
impart a velocity kick of typically tens
to hundreds of km s$^{-1}$ to the resultant merged SMBH \citep{PE62.1, BE73.1}.
The BLR could remain bound to the recoiling SMBH, but
the NLR would remain with the host galaxy \citep{ME06.3,BO07.1}.
A recoiling SMBH could carry with it an accretion disk of
nuclear stars and gas from the BLR, which could power AGN activity for $\sim100$ Myr (e.g.,
\citealt{ME04.3, LO07.1, BO07.1}).  
As a result, recoiling SMBHs could be
visible as AGNs with BLR velocity offsets relative to the
NLR and the host galaxy. A
search of the Sloan Digital Sky Survey for QSOs with such offsets
found no convincing candidates \citep{BO07.1}, and the subsequent
discovery of a Sloan QSO with a 2650 km s$^{-1}$ offset \citep{KO08.2} may in
fact be a SMBH binary \citep{BO09.1, DO08.1} or a superposition of two
AGNs \citep{SH09.1} rather than a recoiling SMBH.
Our offset AGNs are Seyfert 2 galaxies with velocity offsets of the
NLR emission lines, which does not fit the recoiling SMBH
paradigm of a shifted BLR and stationary NLR.

\subsection{Offset and Dual AGNs Are Most Likely \\ the Result of Galaxy Mergers}

We find no significant evidence that the velocity offsets
in our sample are the result of small-scale gas kinematics, AGN
outflows or jets, or recoiling SMBHs, but a combination of gas
kinematics or outflows with an unexpected distribution of dust could
produce the \oiii velocity offsets we measure.  Although some fraction
of the \oiii velocity offsets in our sample
might be caused by gas kinematics or outflows, these effects are
unlikely to explain all of our objects unless the population of AGNs 
at $z \sim 0.7$ is qualitatively different from the well-studied
local population. 

Rather, 
the most plausible explanation for our observations of offset \oiii lines
is an AGN moving with respect to the host galaxy as the result of a
merger.  Inspiralling SMBHs are expected to exist as a consequence of the
well-established evidence for galaxy mergers and galaxies hosting
central SMBHs.  NGC~3341, a
disturbed disk galaxy at $z=0.0271$ with a Seyfert 2 nucleus at a spatial
offset of 5.1 kpc and a blueshifted velocity of 190 km s$^{-1}$
relative to the primary galaxy \citep{BA08.1}, is a local example of a
galaxy merger hosting inspiralling SMBHs. 
Whereas none of the local observations of velocity offsets
we explored in Sections~\ref{gas} and \ref{outflow} would translate into observable velocity offsets at
our redshift range, the \cite{BA08.1} object would.
Because inspiralling SMBHs are an expected consequence of galaxy
mergers and because our offset and dual AGNs are consistent with the
observational signatures expected of inspiralling SMBHs, we conclude
that our observed \oiii velocity offsets are most likely the result of
inspiralling SMBHs in galaxy mergers.

\section{Results and Discussion}

Based on our interpretation that the observed offset and dual AGNs are
the result of mergers, we now use our sample to estimate the fraction
of AGNs hosted by red galaxy mergers and the red galaxy
merger rate.  

\subsection{Evidence for a Link between AGN Activity \\ and Galaxy Mergers}

Combining the extrapolated number of small-velocity-offset AGNs (as
detailed in Section~\ref{low})
 with our 32 definitively detected offset and dual
AGNs, we expect a total of 40 -- 55 offset and dual AGNs in our
data.  Our interpretation of these objects as merger
remnants, therefore, implies that of the 91 red galaxies
harboring AGNs in DEEP2 at 
$0.34 < z < 0.82$, roughly half of the AGNs are moving
relative to their host 
galaxies due to a recent merger.  This striking result, that approximately
half of the 
red galaxies hosting AGNs are also merger remnants, suggests a
strong link between AGN activity and red galaxy mergers.

A number of mechanisms could explain this.  During a galaxy merger
the hot gas of the intergalactic medium is shock heated and cools by
radiation, forming cooling flows that can fuel an AGN.  Numerical
simulations show that mergers between gas-rich, late-type galaxies can
trigger nuclear gas flows that power two separated AGNs that then merge
to form a single central AGN \citep{SP05.2}, and gas-poor, red
galaxy mergers might have enough gas to fuel AGNs in the same way.
Galaxy mergers can also trigger starbursts, and after tens of Myr
some stars evolve into asymptotic giant branch stars, whose
winds might be accreted efficiently onto SMBHs to fuel AGNs
\citep{DA07.2}.

Morphologies based on the imaging of quasar host galaxies also
suggest that $\sim 30\%$ of quasars reside in host galaxies that show
evidence of interactions and mergers \citep{MA03.4, GU06.1}.  More
recently, \cite{UR08.1} found that 85$\%$ of dust-reddened quasars show
evidence of merging in images of their host galaxies.  Our results are
roughly consistent with this finding, though the two samples are very
different.  Our Seyfert 2 AGNs are relatively low luminosity in
comparison to quasars, and the dynamical state of the SMBHs in the
quasar host galaxies is unknown.  We find that AGN activity can be
triggered in merging SMBHs even before coalescence. 

\subsection{Galaxy Merger Rates}

Assuming that all of our offset and dual AGNs are merger remnants, and
making no other assumptions, we set a hard lower limit that at least
2$\%$ of DEEP2 red galaxies at $0.34 < z < 0.82$ have
undergone a merger in the previous $\sim100$ Myr.  Including the
extrapolated number of low-velocity-offset AGNs, as described in Section~\ref{low}, this limit can reach $3\%$.  Our merger fraction can
include minor and major 
mergers, but because we cannot determine the mass ratios of the
progenitor galaxies, we cannot be more specific about the mass range probed. 

An estimate of the galaxy merger rate depends on both the timescale
over which two 
SMBHs merge and the fraction of SMBHs that are visible as AGNs.
Recall that two SMBHs in a merger spend $\mathrm{t_{combine}} \sim100$ Myr at
separations $\gtrsim 1$ kpc.  At separations $\lesssim 1$ kpc,
the SMBHs quickly sink to the bottom of the galaxy's potential well
due to dynamical friction and as a close binary can no longer be kinematically
distinguished because their separation is smaller than the size of the
\oiii emitting region.
The fraction of SMBHs that power AGNs is $\mathrm{f_{lum}} \sim 10\%$
\citep{MO09.2}, roughly consistent with our measured fraction of
red galaxies hosting AGNs ($5\%$) and the fraction of dual AGNs
in our sample of offset and dual AGNs ($6\%$).

We include the expected number of offset and dual AGNs at low velocity
differences to estimate the merger rate from our data.
Correcting the parent sample of 1881 red galaxies by the fraction for
which we can determine good absorption redshifts (95$\%$; the rest are
removed because the continuum signal-to-noise ratio is too low), we
convert our estimate of 40 -- 55 offset and dual AGNs into a galaxy
merger fraction 
of $22$ -- $31\%$ (10 $\%$ / $\mathrm{f_{lum}}$), or a galaxy merger rate
of $2.2$ -- $3.1$ mergers Gyr$^{-1}$ (100 Myr / 
$\mathrm{t_{combine}}$) (10 $\%$ / $\mathrm{f_{lum}}$) for DEEP2
red galaxies at $0.34 < z < 0.82$.  Again, this merger rate can
include both minor and major mergers.  

The merger fraction of all galaxies is generally parameterized to evolve 
as $(1+z)^n$, where measurements
of the exponent $n$ range from 0 -- 4 (e.g., \citealt{YE95.1,WO95.1,PA97.1}).  If
the red galaxy merger 
fraction evolves significantly, we 
would expect a higher merger fraction for our sample than for a local
sample.  However, large uncertainties in the merger fraction evolution with
redshift prevent us from making any direct comparisons. 

A rough comparison, however, can be made between our merger fraction and other galaxy merger fractions derived observationally and semianalytically.
\cite{VA05.1} found
that 35$\%$ of early-type galaxies at $0.05 < z < 0.2$ exhibit
morphological signs of a recent merger, such as tidal 
tails and asymmetries in surface brightness.
In addition, counts of close dynamical pairs of galaxies suggest that
24$\%$ of present day
red galaxies have experienced gas-poor mergers with
luminosity ratios 
between 1:4 and 4:1 since $z \sim 1$ \citep{LI08.2}.  Finally,
semianalytic models of gas-poor major mergers suggest that $\sim 5\%$
of $M_B \lesssim -20$ galaxies at $0.1 < z < 1.1$ have had a major
merger in the past 1 Gyr \citep{BE06.3}. 

Our red galaxy merger fraction of $\sim30\%$ is roughly
consistent with galaxy merger fractions estimated from galaxy
morphologies and close pairs of galaxies.  This agreement is further
evidence that the AGN velocity offsets we measure are the result of
inspiralling SMBHs in galaxy mergers and not AGN outflows or gas
kinematics (see Section~\ref{interpret}).

Our merger fraction is much
higher than the $\sim 5\%$ merger fraction derived by semianalytic
models, but this difference can be explained by differences in the
merger mass ratios.  The merger fraction from semianalytics considers
only major mergers (mass ratios of 1:1 to 4:1), while our
technique is sensitive to much higher mass ratios, provided that the
less massive progenitor galaxy hosts a SMBH that can power an AGN.
Significant differences in the methods and assumptions of different
observational and theoretical approaches -- especially the timescales involved -- prevent
a more direct comparison between merger fractions. 
       
\section{Conclusions}
 
We have searched the DEEP2 Galaxy Redshift Survey for AGNs with
velocities significantly different from the mean
of the host galaxy's 
stars, and have identified 30 objects with one set of offset AGN-fueled
emission lines (``offset AGNs'') and two objects with two spatially
resolved sets of AGN emission lines (``dual AGNs'').  
Although a conspiracy between dust and gas kinematics or outflows
could cause \oiii velocity offsets like those observed in our sample,
our entire sample is unlikely to be the result of such effects
unless the population of AGNs 
at our redshift range $0.34 < z < 0.82$ is qualitatively different
from the well-studied local population. 
Rather, the more plausible interpretation of our sample is that they
are the results of recent galaxy mergers, during which
two SMBHs spiral to the remnant's center at velocities
different from the mean of the host galaxy's stars.  If one or both of
the SMBHs power AGNs, they will appear in our sample.  Based on this
interpretation, our main results are as follows.

1. We present the first systematic search for inspiralling SMBHs in
galaxy mergers, from which we identify 32 such objects: 30 offset AGNs
and two dual AGNs.  Our 
technique of selecting these objects by the velocity offsets of
their AGN emission lines relative to the host galaxy's stars is a new
and powerful way of identifying galaxies that are the products of
recent mergers. 

2. About half of the red galaxies hosting AGNs are also merger
remnants, which signals a strong correlation between AGN activity and
mergers of gas-poor galaxies.

3. For DEEP2 red galaxies at $0.34 < z < 0.82$, the merger fraction is $\sim 30\%$ (10 $\%$ / $\mathrm{f_{lum}}$),
with a hard lower limit of $2\%$, and the merger rate is $\sim3$
mergers Gyr$^{-1}$ (100 Myr / $\mathrm{t_{combine}}$) (10 $\%$ /
$\mathrm{f_{lum}}$). 
This merger rate can include both minor and major mergers.  Our merger
fraction is in agreement with merger fractions derived through other
observational techniques, lending additional support to our interpretation
that the AGN velocity offsets we measure are the result of
inspiralling SMBHs in galaxy mergers and not other physical mechanisms.

While other observational methods of determining the galaxy merger
rate include no information about the state of the SMBHs in the
merger, our new method is unique in that it selects merger-remnant galaxies
with inspiralling SMBHs.  Merging SMBHs are predicted to produce
gravity waves, and although our offset and dual AGNs do not probe SMBH
separations on the small scales where gravitational radiation is
significant, our measurements also constrain the rate of SMBH mergers
of interest to proposed gravitational wave detectors such as LISA
\citep{BE98.2}. 
         
\acknowledgements J.M.C. acknowledges support from NSF grant
AST-0507428.  B.F.G. was supported by the U.S. Department of Energy
under contract number DE-AC3-76SF00515.  M.C.C. acknowledges support by
NASA through the Spitzer Space Telescope Fellowship program. A.L.C. is
supported by NASA 
through Hubble Fellowship grants HF-01182.01-A, awarded by the Space
Telescope Science Institute.  This work was also supported 
in part by NSF grants AST-0071048, AST-0071198, and AST-0507483. B.F.G. thanks the Aspen Center
for Physics for its hospitality, and we also thank Jay Dunn and Mike Crenshaw for illuminating
conversations.

\bibliographystyle{apj}
\bibliography{comerford_smbh}

\end{document}